\documentclass[11pt]{iopart}
\usepackage{epstopdf}
\usepackage[dvipdfm]{graphicx}
\usepackage{tabularx}
\usepackage{color}
\usepackage{amssymb,amsthm}

\def\eV2{eV$^{2}$}

\renewcommand\footnotemark{}

\begin{document}

\title{
 NESSiE: The Experimental Sterile Neutrino Search in Short-Base-Line at CERN}
\footnotetext{Prensented at the Lake Louise Winter 2013 Conference, Banff, Alberta, Canada, 17-23 February 2013.}

\author{UMUT KOSE\\On behalf of NESSiE Collaboration}
\address{ 
  INFN Sezione di Padova,
  \\
  I-35131 Padova, Italy\\
}
 
\eads{\mailto{umut.kose@cern.ch}}



\begin{abstract}

Several different experimental results are indicating the existence of anomalies in the neutrino sector.
Models beyond the standard model have been developed to explain these results 
and involve one or more additional neutrinos that do not weakly interact. 
A new experimental program is therefore needed to study this potential 
new physics with a possibly new Short-Base-Line neutrino beam at CERN. 
 CERN is actually promoting the start up of a New Neutrino Facility in the North Area site, 
 which may host  two complementary detectors, one based on LAr technology and one 
 corresponding to a muon spectrometer. The system is doubled in two different sites.
 With regards to the latter option,
 NESSiE, Neutrino Experiment with Spectrometers in Europe, 
 had been proposed for the search of sterile neutrinos 
  studying Charged Current (CC) muon neutrino and antineutrino ineractions. 
  The detectors consists of two magnetic  spectrometers to be located in two sites: 
  "Near" and "Far" from the proton target of the CERN-SPS beam. 
  Each spectrometer  will be complemented by an ICARUS-like LAr target in order 
  to allow also Neutral Current (NC) and electron neutrino CC interactions  reconstruction.
\end{abstract}

\normalsize\baselineskip=15pt

\section{Introduction to sterile neutrino}

Most of existing data on neutrino oscillations from the solar~\cite{ref:solar}, atmospheric~\cite{ref:atmospheric}, reactor~\cite{ref:reactor} and accelerator~\cite{ref:accelerator} experiments
have established a framework of neutrino oscillations among three flavor neutrinos mixed with three mass eigenstates.
These sets of eigenstates are related through a $3\times3$ unitary matrix, called the Pontecorvo-Maki-Nakagawa-Sakata (PMNS) matrix~\cite{ref:PMNS} which 
is commonly parameterized by three angle, $\theta_{ij}$, and a CP-violating phase, $\delta$.
The oscillation parameters from the global fits of the current neutrino oscillation data with 1 $\sigma$ uncertainity have been determined
as $\Delta{m}_{21}^{2}\simeq 7.54^{+0.26}_{-0.22}\times10^{-5}eV^{2},  
\Delta{m}_{31}^{2}\simeq\Delta{m}_{32}^{2}\simeq 2.43^{+0.06}_{-0.10}\times10^{-3}eV^{2}, 
\sin^{2}\theta_{12}\simeq 0.307^{+0.18}_{-0.16}, 
\sin^{2}\theta_{23}\simeq 0.386^{+0.24}_{-0.21}$ 
and 
$\sin^{2}\theta_{13}\simeq 0.0241\pm-0.0025$ for normal hierarchy~\cite{ref:pdg,ref:fogli}. CP violation phases are unknown.

On the other hand, there exist a few experimental results, at the level of anomalies (i.e. with a significance around 2-4 $\sigma$), 
that cannot be explained in the standard three flavour picture.
The first anomaly is coming from the LSND experiment~\cite{ref:lsnd}. They studied the transitions 
$\bar\nu_\mu\to\bar\nu_e$ with a baseline of $L/E \sim 1(m/MeV)$, where $E$ ($\sim 30$ MeV) is the neutrino energy
 and $L$ ($\sim 30$ m) is the distance between source and detector. They reported a $\bar{\nu}_{e}$ excess of about 3.8 $\sigma$
above the expected background including standard three flavour neutrino oscilations (``LSND anomaly''). 
This excess requires $\bar\nu_\mu\to\bar\nu_e$ oscillations with $\Delta{m}^{2}$ in the range from 0.2 \eV2 to 2 \eV2.
If neutrino oscillations are responsible, a solution might require additional, sterile, neutrino species which do not couple to the Z boson.

The MiniBooNE experiment~\cite{ref:miniboone} was designed to examine the LSND parameter space at the same baseline $L/E$ by studying
$\nu_{\mu}\to\nu_{e}$ and $\bar\nu_\mu\to\bar\nu_e$ transitions. From combined analysis of both channels,
an excess of 3.8 $\sigma$ in the range $200< E_{\nu} < 1250 $ MeV have been observed (``MiniBooNE anomaly''). 
By interpreting the data in terms of neutrino oscillations,
the extracted parameter values are consistent with the ones coming from LSND.

Re-evaluation of the neutrino flux emitted by nuclear reactors \cite{ref:reactorflux} has been increased by $\sim3.5\%$.
Based on the new flux calculation, the results of previous short-baseline ($L \lesssim 100$~m) reactor experiments
show a $\sim6\%$ deficit (about $\sim$3 $\sigma$ effect) in the measured $\bar{\nu}_{e}$ flux. This deficit can be explained by assuming 
$\bar{\nu}_e$ disappearance due to oscillations with $\Delta m^2 \sim1$~eV$^2$ (``reactor anomaly'').

An additional anomaly was raised from radioactive source experiments at the Gallium solar neutrino
experiments SAGE~\cite{ref:sage} and GALLEX~\cite{ref:gallex}.
They have obtained an event rate induced by $\nu_{e}$ fluxes produced by intense $^{51}$Cr and $^{37}$Ar sources which is lower of about
$\sim 15\%$ than expected. This effect can be explained by the hypothesis of $\nu_e$ disappearance due to oscillations with $\Delta m^2 \gtrsim
1$~eV$^2$~\cite{ref:giunti_laveder_gallium} (``Gallium anomaly'').

These anomalies in neutrino oscillation data can be explained by a hypothetical 
fourth neutrino separated from the three standard neutrinos by a squared mass difference of few \eV2. 
Furthermore, analysis of cosmological data~\cite{ref:cosmology} such as Cosmic Microwave Background, Big Bang Nucleosynthesis allows one extra
sterile neutrino.

Therefore we would conclude that these studies are becoming one of the most important topics to be addressed in neutrino physics.
A large number of experiments will be coming up with the aim of investigating their possible 
presence with a variety of methods and approaches~\cite{ref:whitepaper}. 

One of them is the ICARUS-NESSiE experiment (SPSC-P-347)~\cite{ref:icarus_nessie}. It is a joint proposal for the search of sterile neutrinos with a 
short-baseline neutrino beam based at CERN. In the following, the new short-baseline neutrino beam facility at CERN and the NESSiE detector concept, as well as the physics
reach of the experiment, will be discussed. The detail on ICARUS LAr-TPC detector and techniques have been discussed elsewhere~\cite{ref:icarus}.

\section{CERN Neutrino Facility, CENF}
\begin{figure}[htbp]
\begin{center}
\includegraphics[height=5cm,width=12cm]{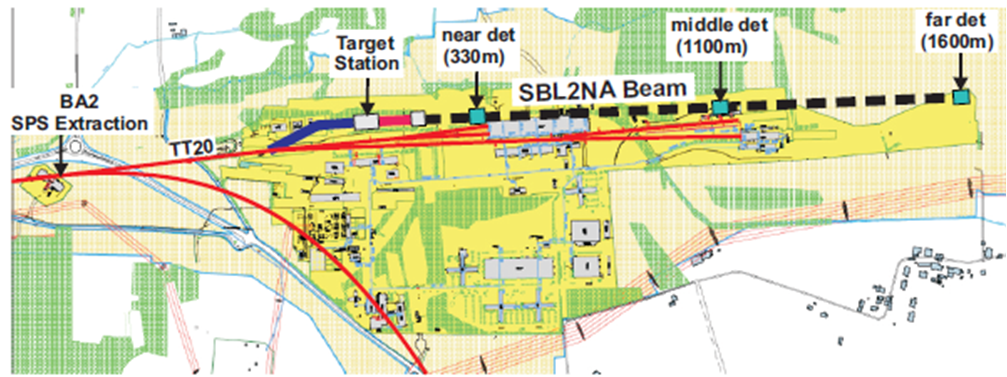}
\end{center}
    \caption{The new SPS North Area neutrino beam layout. Main parameters are: primary beam: 100 GeV; fast extracted from SPS; 
    target station next to TCC2, $\sim$11~m underground; decay pipe: 110~m, 3~m diameter; beam dump: 15~m of Fe with graphite core, 
    followed by muon stations; neutrino beam angle: pointing upwards; at $\sim$3 m in the far detector $\sim$5~mrad slope.}
    \label{fig:CENF}
\end{figure}

A new short-baseline neutrino beam facility, CENF~\cite{ref:CENF}, was proposed in the CERN North Area, shown in Figure~\ref{fig:CENF}, in order to
study anomalies mentioned above.  It will consist of an SPS fast extraction system and a new proton transfer line to bring high energy protons
to the target area. The primary target will be followed by a decay tunnel terminated by a beam dump.

The neutrino beam will be produced by accelerating protons of
100 GeV/c by the CERN Super Synchrotron (SPS). These protons will be ejected towards a
graphite neutrino production target in two extractions, separated in time by 50 ms. 
Each SPS cycle length will be 3.6 s long.
Secondary particles, mainly charged pions and kaons, will be focused by magnetic horns and decay in flight into neutrino in
 110 m long hellium filled decay tunnel of about 3 m diameter. Neutrino(antineutrino) beam with a peak energy around 2 GeV, will travel through the identical detectors located in "Near" and "Far" detector
 site, 450 m and 1600 m, respectively. The proton beam intensity of $4.5\times10^{19}$ pot/year is expected. At the appropriate oscillation path L/E$_{\nu}$, the experiment is going to
 undoubtedly shed more light on the $\Delta{m}^2$ window for expected anomalies.

\section{The NESSiE Detectors}

\begin{figure}[htb]
\begin{center}
\includegraphics[height=5cm,width=8cm]{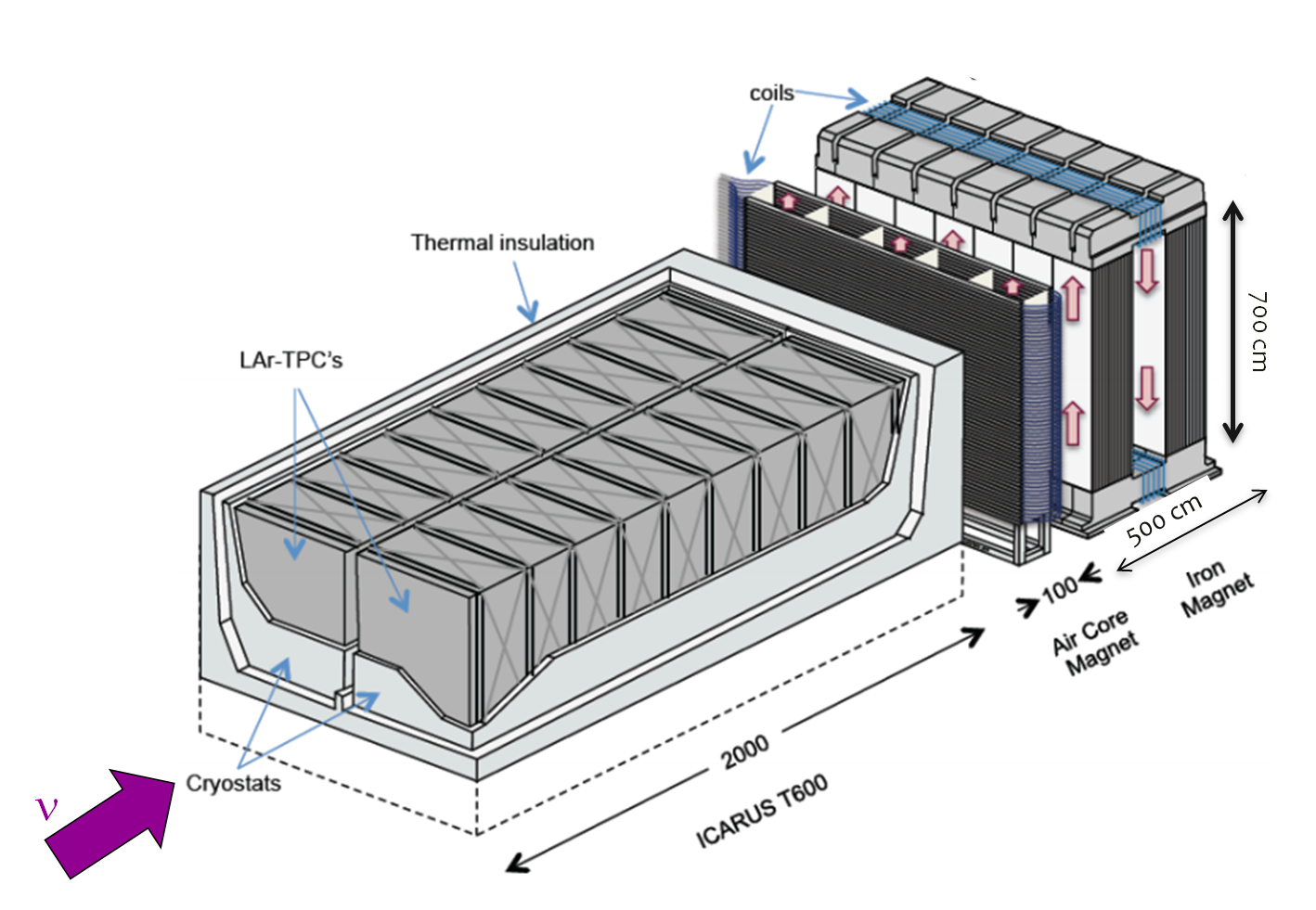}
\end{center}
\caption{
Sketch of the ICARUS-NESSiE far detectors.
}\label{fig:detector}
\end{figure}

The NESSiE Near and Far detectors, as shown in Figure~\ref{fig:detector},  
will be placed just downstream of ICARUS LAr-TPC detector, in order to 
measure with high precision the charge and the momentum of muons produced
by neutrino interactions in the LAr target and those interacting in the spectrometer itself.
The NESSiE detector will consist of an air-core magnet (ACM) followed by an iron-core magnet (ICM). 
The ICM is dedicated to the precise reconstruction of high-energy muons
(up to 30 GeV) and to reach few $\%$ precision, through range measurement, on the momentum of
muons with energy lower than 4 GeV.  The ACM covers the low momentum region where it
ensures high momentum resolution and charge discrimination 
(allowing to separately study the $\nu$ and $\bar\nu$ component). 

The dipolar ICM spectrometers are instrumented with vertical iron plates (210 slabs  with a total of 800 tons for the Near,
294 iron slabs with a total of 1500 tons for the Far detector) interleaved
with detector layers composed by Resistive Plate Chambers (RPCs) for a total of 700 m$^2$ and 12000 digital channels in the Near and 
1800 m$^2$ and 20000 digital channels in the Far sites. RPCs provide the tracking inside the magnet with 1 cm resolution
and range measurement for stopping muons. The typical values of the magnetic field is about 1.5 T.
Most of the detectors from the OPERA spectrometers~\cite{ref:operadetector} might be recovered and re-used. 
The possibility of reusing the iron slabs of the OPERA spectrometers is under study, too.

\begin{figure}[htb]
\begin{center}
  \includegraphics[width=0.45\textwidth]{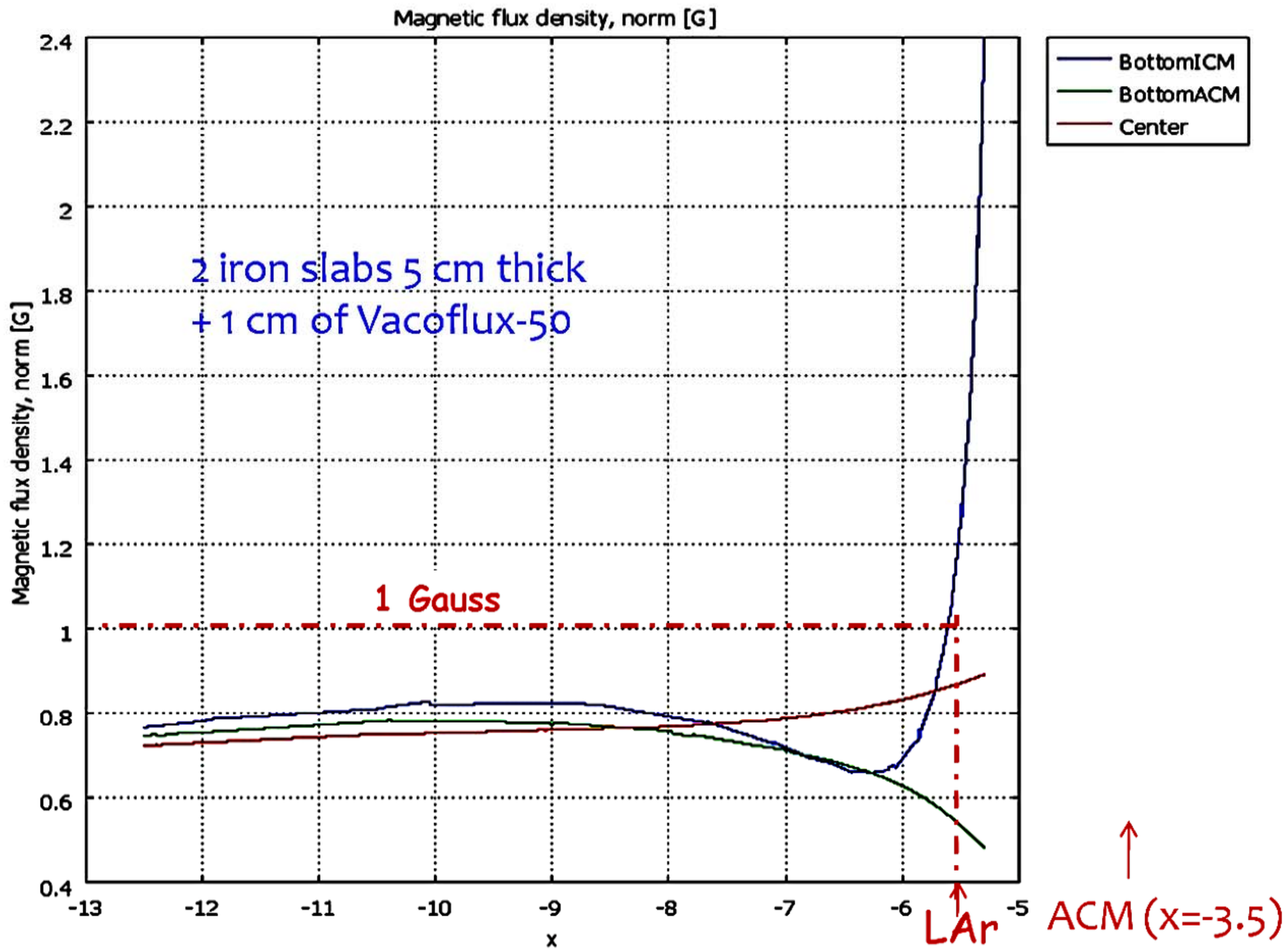}
\hspace{0.5cm}
\includegraphics[width=0.45\textwidth]{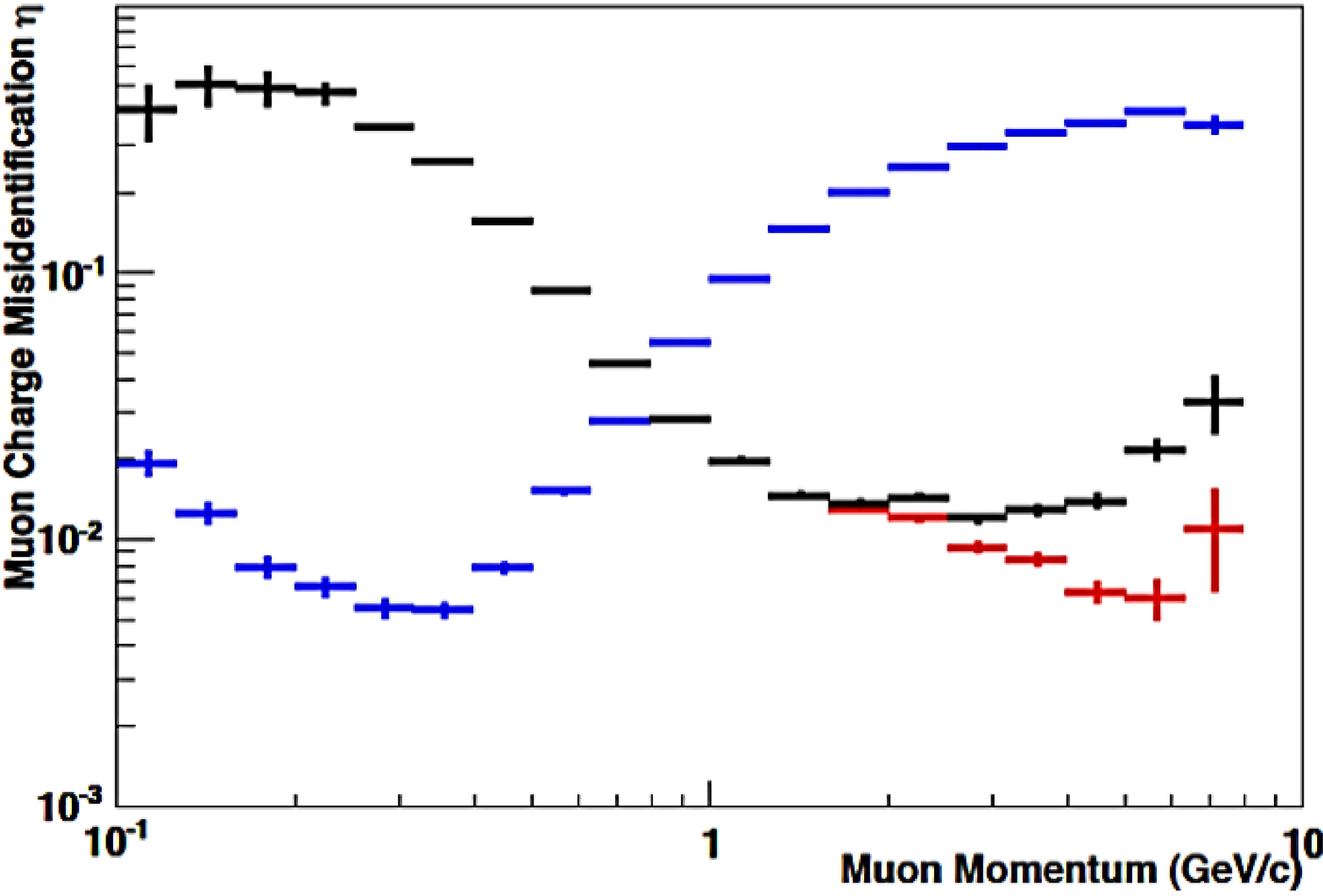}
\end{center}
\caption{
Left: Fringe field outside the ACM magnet, Right: The charge mis-idenfication
percentage including all selection, efficiency and reconstrcution procedures by the NESSiE
system. Blue dots correspond to the measurement performed by ACM, the red (black)
dots correspond to the one by the ICM with the two (one) arms.
}
\label{fig:fringe}
\end{figure}

The ACM is instead a new design, using 51 (39) coils 9 meters long in the straight parts and two half circular
bending regions for the return of the conductors outside the beam region in the Far (Near) site. 
Aluminium material have been chosen both for the 
conducting cables and the supporting structure. All coils are connected electrically and hydraulically in series. Planes of
High Precision Tracker (HPT) will be placed in the ACM in order to provide tracking with 1 mm resolution. The total mass of the ACM
is about 6 tons. The magnetic field in ACM can reach 0.1 T using a dedicated power supply. 
And the fringe field outside the magnet is below the constraints imposed by the LAr electronics, as shown in Figure~\ref{fig:fringe}.

Both ICM and ACM are going to provide a charge misidentification probability 
as low as 1\% over a momentum range from 0.1 to 10 GeV, 
as shown in Fig.~\ref{fig:fringe}.

\section{Expected results}

Assuming the $\Delta{m^2}$ around 2 \eV2 and $4.5\times10^{19}$ pots for 1 year of operation, 
either with negative or positive polarity beam,
the expected CC interaction rates in the LAr-TPCs at the Near (effective 119 t) and Far
locations 
(effective 476 t), and the expected rates of fully reconstructed events in the
NESSiE spectrometers at the Near (effective 241 t) and Far locations (effective 661 t),
with and without LAr contribution are shown in Table~\ref{tab:rates}. The spectrometer will be able to correctly identify
about 40\% of all the CC events produced in, and escaped from the LAr-TPC's, both in the near
and far sites. Therefore complete measurement of the CC event spectra will be possible,
along with the NC/CC event ratio in synergy with the LAr-TPC and the relative background systematics.
\begin{table}[htbp!]
  \begin{center}
  \begin{tabular}{lcccc}
  \hline
  & NEAR                            & NEAR             & FAR                        & FAR            \\
  &       (Negative foc.)           & (Positive foc.)  & (Negative foc.)            & (Positive foc.) \\
  \hline
  $\nu_{e}+\bar{\nu}_{e}$(LAr) & 35K & 54K & 4.2K & 6.4K \\
  $\nu_{\mu}+\bar{\nu}_{\mu}$(LAr) & 2000K & 5200K & 270K & 670K \\
  Appearance Test Point & 590 & 1900 & 360 & 910 \\
  $\nu_{\mu}$ CC (NESSiE$+$LAr) & 230 K & 1200 K & 21 K & 110 K \\
  $\nu_{\mu}$ CC (NESSiE alone) & 1150 K & 3600 K & 94 K & 280 K \\
  $\overline{\nu}_{\mu}$ CC (NESSiE$+$LAr) & 370 K & 56 K & 33 K & 6.9 K \\
  $\overline{\nu}_{\mu}$ CC (NESSiE alone) & 1100 K & 300 K & 89 K & 22 K \\
 Disappearance Test Point & 1800 & 4700 & 1700 & 5000 \\
 \hline
  \end{tabular}
  \end{center}
    \caption{The expected rates of interaction (LAr) and reconstructed (NESSiE) events 1 year of operation. Values for 
  $\Delta m^2$ around 2 eV$^2$ are reported as example.}
  \label{tab:rates}
\end{table}
On the top of that a high number of CC events will be produced in the spectrometers. This will also 
allow to study the NC/CC ratio in an extended energy range, and to perform an independent measure of $\nu_\mu$\ disappearance

\begin{figure}[htbp]
\begin{center}
  \includegraphics[width=0.45\textwidth]{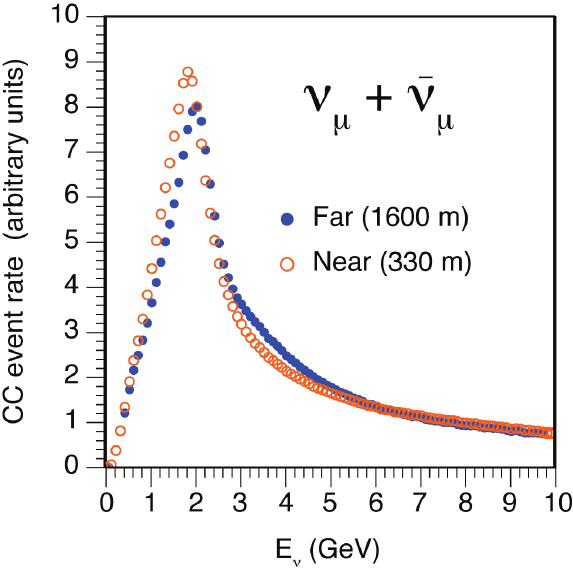}
\hspace{0.5cm}
  \includegraphics[width=0.45\textwidth,height=0.34\textheight]{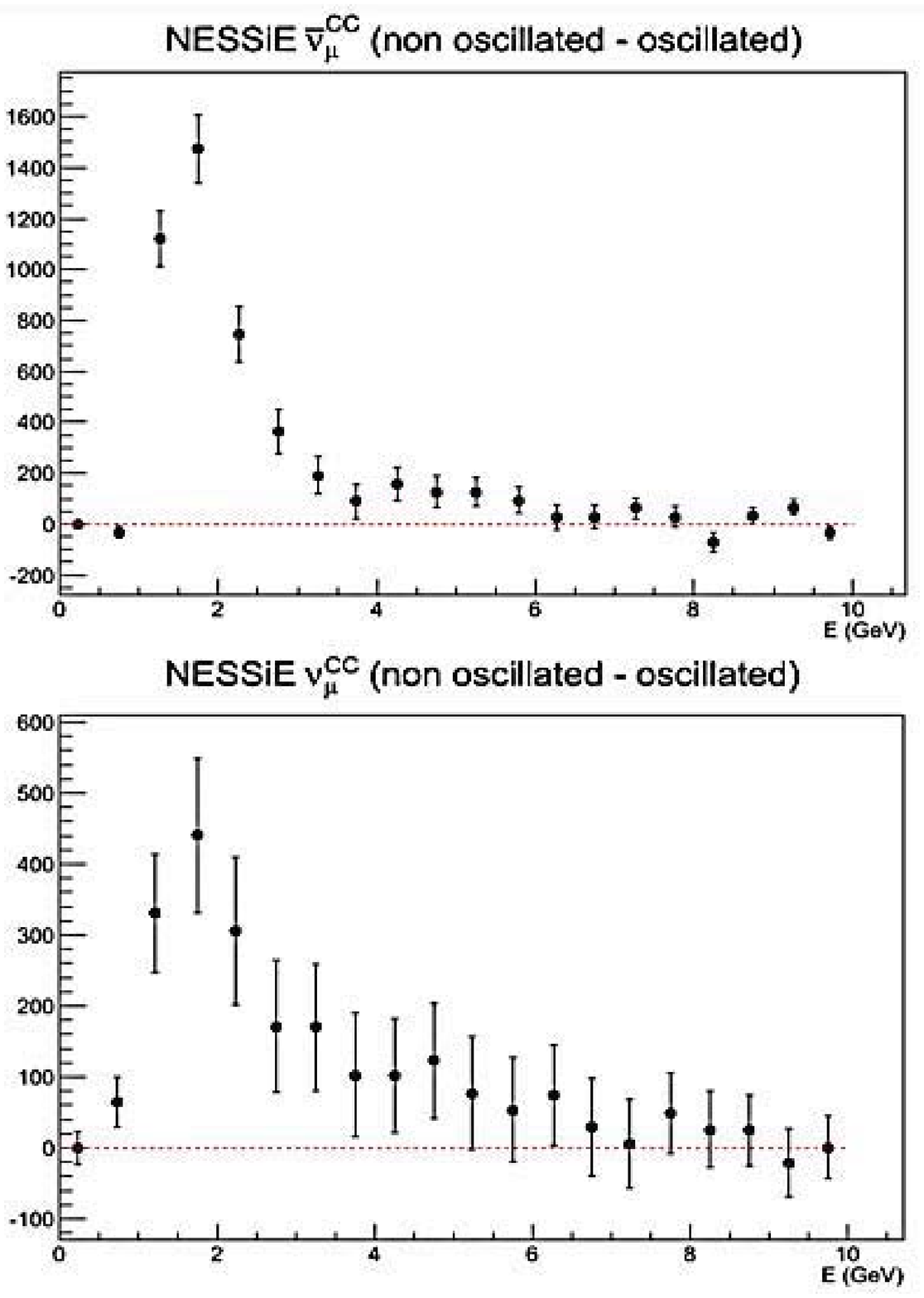}
    \caption{Left: Muon neutrino CC interaction spectra, at Near and Far positions, Right: Difference between rates
    estimated with and without oscillation, Top: antineutrino CC events and Bottom: neutrino CC events.}
    \label{fig:rates}
\end{center}
\end{figure}

The shapes of the radial and energy spectra of the neutrino beam component, in the Far and Near locations,
are practically identical. In the absence of oscillations, all cross sections and experimental biases cancel out,
and two experimentally observed event distributions must be identical.
Any emerged difference of the event distributions at the locations of the two detector might be attributed to 
the possible existence of neutrino oscillation due to additional sterile neutrinos. 
The difference of the expected spectra for the measured CC muon events for non-oscillation and with
 the oscillation hypothesis, for both neutrino and antineutrino exposure, are shown in Figure~\ref{fig:rates}, for the "NESSiE alone"
 detection.

\begin{figure}[htbp]
\centering
\includegraphics[width=0.55\textwidth]{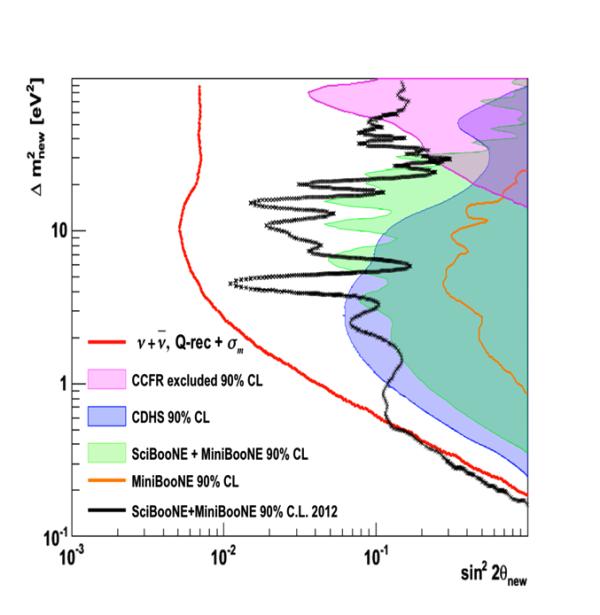}
\caption{
	Sensitivity plot (at 90\% C.L.) considering 3 years of the CERN neutrino beam (2 years in antineutrino
	and 1 year in neutrino mode) from CC events fully reconstructed in NESSiE+LAr. Red line: $\nu_\mu$ from 
	CCFR~\cite{ref:ccfr}, CDHS~\cite{ref:cdhs} and SciBooNE+MiniBooNE~\cite{ref:sciboone} experiments (at 90\% C.L.).
	Orange line: recent exclusion limits on $\nu_\mu$ from MiniBooNE alone measurement~\cite{ref:mini}.
}
\label{fig:limits}
\end{figure}

The $\nu_{\mu}$ disappearance signal is well studied by the NESSiE spectrometers, with large
statistics and by disentangling $\nu_{\mu}$ from $\bar{\nu}_{\mu}$~\cite{ref:icarus_nessie}. 
As an example, Figure~\ref{fig:limits} shows the
sensitivity plot (at $90\%$ C.L.) for two years negative-focusing (neutrino) plus one year positive
focusing (antineutrino). A large extension of the present limits for $\nu_{\mu}$ by CDHS~\cite{ref:cdhs} and the recent
SciBooNE+MiniBooNE~\cite{ref:sciboone} will be achievable in the sin$^{2}2\theta$, and $\Delta{m^2}$ space.

The physics reach on electron neutrino oscillation both in appearance and disappearance mode in ICARUS-NESSiE
experiment can be found elsewhere~\cite{ref:icarus_nessie}.

\section{Conclusions}
The ICARUS-NESSiE experiment will explore in a definitive way a region of
parameter space completely covering the possible anomaly claimed by LSND and a large 
fraction of the region relevant to the reactor anomalies. Looking for 
the muon neutrino CC and NC disappearance as well as the electron neutrino oscillation both in appearance and disappearance
modes will allow disentangling the different possible couplings to sterile neutrinos and cover all possible 
light sterile neutrino signatures for masses up to a few \eV2.

The measurements of the neutrino flux at the Near detector in the full muon momentum range
is relevant to keep the systematic errors at the lowest possible values. Moreover, the measurement of the muon charge
will enable to separate $\nu_\mu$ from $\bar{\nu}_\mu$ which is very important since the $\nu_\mu$ contamination
is large in antineutrino beam mode. This will also allow to fully exploit the experimental capability
of observing any difference between $\nu_{\mu}\to\nu_{e}$ and $\bar{\nu}_{\mu}\to\bar{\nu}_{e}$ (CP violation signature).

\title

{\small References:}

\begin{thebibliography}{100}
\bibitem{ref:solar}
 K. S. Hirata {\it et al.} [Kamiokande-II Collaboration], 
 Phys.\ Rev.\ Lett.\ {\bf 63}, 16 (1989); \\
 J. Abdurashitov et al. [SAGE Collaboration], 
 Phys.\ Lett.\ B {\bf 328}, 234 (1994); \\
 P. Anselmann {\it et al.} [GALLEX Collaboration], 
 Phys.\ Lett.\ B {\bf 327}, 377 (1994); \\
 S. Fukuda {\it et al.} [Super-Kamiokande Collaboration],
 Phys.\ Rev.\ Lett.\ {\bf 86}, 5651 (2001); \\
 Q. R. Ahmad {\it et al.} [SNO Collaboration], 
 Phys.\ Rev.\ Lett.\ {\bf 89}, 011301 (2002); \\
 C. Arpesella {\it et al.} [Borexino Collaboration], 
 Phys.\ Rev.\ Lett.\ {\bf 101}, 091302 (2008),
\bibitem{ref:atmospheric}
 Y. Fukuda {\it et al.} [Super-Kamiokande Collaboration],
 Phys.\ Lett.\ B {\bf 436}, 33 (1998), arXiv:hep-ex/9805006 [hep-ex];\\
 K. Hirata {\it et al.} [Kamiokande-II Collaboration], 
 Phys.\ Lett.\ B {\bf 205}, 416 (1988);\\
 R. Becker-Szendy {\it et al.} [IMB Collaboration], 
 Phys.\ Rev.\ D {\bf 46}, 3720 (1992);\\
 S. P. Ahlen {\it et al.} [MACRO Collaboration], 
 Phys.\ Lett.\ B {\bf 357}, 481 (1995);\\
 W. Allison {\it et al.} , 
 Phys.\ Lett.  B {\bf 391}, 491 (1997), arXiv:hep-ex/9611007 [hep-ex];\\
 R. Wendell et al. [Super-Kamiokande Collaboration],
 Phys.\ Rev.\ D {\bf 81}, 092004 (2010),

\bibitem{ref:reactor}
 F. P. An {\it et al.} [Daya Bay Collaboration], 
 Phys.\ Rev.\ Lett.\ {\it 108}, 171803 (2012);\\
 J. K. Ahn {\it et al.} [RENO Collaboration], 
 Phys.\ Rev.\ Lett.\ {\it 108}, 191802 (2012);\\
 M. Apollonio {\it et al.} [CHOOZ Collaboration], 
 Eur. Phys. J. C {\it 27}, 331 (2003);\\
 B. Armbruster {\it et al.} [KARMEN Collaboration], 
 Phys.\ Rev.\ D {\bf 65} 112001 (2002), [hep-ex/0203021],
\bibitem{ref:accelerator}
 M. H. Ahn {\it et al.} [K2K Collaboration], 
 Phys.\ Rev.\ D {\bf 74}, 072003 (2006);\\
 P. Adamson {\it et al.} [MINOS Collaboration], 
 Phys.\ Rev.\ Lett.\ 108, 191801 (2012)\\
 K. Abe {\it et al.} [T2K Collaboration], 
 Phys.\ Rev.\ D {\bf 85}, 031103 (2012);\\
 N. Agafonova {\it et al.} [OPERA Collaboration], 
 Phys.\ Lett.\ B {\bf 691}, 138 (2010).

\bibitem{ref:PMNS} B. Pontecorvo,
  {\it JETP}{\bf 34}, 172 (1958); \\
  V. N. Gribov and B. Pontecorvo,
  {\it Phys. Lett. B} {\bf 28}, 493 (1969); \\
  Z. Maki, M. Nakagawa and S. Sakata,
  {\it Prog. Theor. Phys.} {\bf 28}, 870 (1962).
\bibitem{ref:pdg}
  J.~Beringer {\it et al.}  [Particle Data Group Collaboration],
  Phys.\ Rev.\ D {\bf 86}, 010001 (2012). 
\bibitem{ref:fogli}
  G. L Fogli {\it et al.},
  Phys.\ Rev.\ D {\bf 86}, 013012 (2012);\\
  E. Lisi, presentation given in NEUTEL 2013, 11-15 March 2013, Venice, Italy.  
\bibitem{ref:lsnd}
  A.~Aguilar-Arevalo {\it et al.}  [LSND Collaboration],
  Phys.\ Rev.\ D {\bf 64}, 112007 (2001)  
\bibitem{ref:miniboone}
  A.~A.~Aguilar-Arevalo {\it et al.}  [MiniBooNE Collaboration],
  Phys.\ Rev.\ Lett.\  {\bf 98}, 231801 (2007)  
 A.~A.~Aguilar-Arevalo {\it et al.}  [MiniBooNE Collaboration],
  Phys.\ Rev.\ Lett.\  {\bf 105}, 181801 (2010)  
  A.~A.~Aguilar-Arevalo {\it et al.}  [MiniBooNE Collaboration],
  arXiv:1207.4809 [hep-ex].
\bibitem{ref:reactorflux}
 G.~Mention {\it et al.}
  Phys.\ Rev.\ D {\bf 83}, 073006 (2011)  
\bibitem{ref:sage}
  J.~N.~Abdurashitov {\it et al.} [SAGE Collaboration],
  Phys.\ Rev.\ C {\bf 59}, 2246 (1999)  [arXiv:hep-ph/9803418 [hep-ph]];\\
  Phys.\ Rev.\ C {\bf 80}, 015807 (2009)  [arXiv:0901.2200 [nucl-ex]].
\bibitem{ref:gallex}
  W.~Hampel {\it et al.}  [GALLEX Collaboration],
  Phys.\ Lett.\ B {\bf 420}, 114 (1998).
\bibitem{ref:giunti_laveder_gallium}
  M. A. Acero, C. Giunti and M. Laveder, 
  Phys.\ Rev.\ D {\bf 78}, 073009 (2008)  [arXiv:0711.4222],
  C.~Giunti and M.~Laveder,
  Phys.\ Rev.\ C {\bf 83}, 065504 (2011)  [arXiv:1006.3244].
\bibitem{ref:cosmology} 
  E. Komatsu {\it et al.}, [WMAP Collaboration], 
  Astrophys.\ J.\ Suppl. {\bf 192}, 18 (2011) [arXiv:1001.4538 [astro-ph.CO]]; \\
  P. A. R. Ade {\it et al.}, [Planck Collaboration], ``Plack 2013 results. XVI'', arXiv:1303.5076 [astro-ph.CO];\\
  G. Mangano and P.D. Serpico, 
  Phys.\ Lett.\ B {\bf 701}, 296 (2011).
\bibitem{ref:whitepaper}
  K.~N.~Abazajian {\it et al.},``Light Sterile Neutrinos: A White Paper,''
  arXiv:1204.5379 [hep-ph].  
\bibitem{ref:icarus_nessie}
  P. Bernardini {\it et al.} [NESSiE Collaboration] SPSC-P-343, arXiv:1111.2242v1 (2011);\\
  M. Antonello {\it et al.} [ICARUS-NESSiE Collaboration], SPSC-P-347 (2012), arXiv:1203.3432 (2012), arXiv:1208.0862v2
(2012),
\bibitem{ref:icarus}
  C. Rubbia {\it et al.}, [ICARUS Collaboration],
  JINST {\bf 6}, (2011) P07011;\\
  A. Ankowski {\it et al.}, [ICARUS Collaboration], 
  Acta\ Phys.\ Polon.\ {\bf B41}, 103 (2010),
\bibitem{ref:CENF}
{\it Letter of Intent for the new CERN Neutrino Facility (CENF)},\\
https://edms.cern.ch/file/1260766/1/LoI.pdf.
\bibitem{ref:operadetector}
  Acquafredda R. {\it et al.} [OPERA Collaboration],
  JINST {\bf 4}, (2009), P04018,
\bibitem{ref:ccfr} 
  Stockdale I. E. {\it et al.} [CCFR Collaboration ],
  Phys.\ Rev.\ Lett.\ {\bf 52}, 1384 (1984),
 
\bibitem{ref:cdhs} 
  Dydak F. {\it et al.} [CDHSW Collaboration],
  Phys.\ Lett.\ B {\bf 134}, 281 (1984),
\bibitem{ref:sciboone}
  Mahn K. B. M. {\it et al.} [MiniBooNE and SciBooNE Collaborations], 
  Phys.\ Rev.\ D\ {\bf 85}, 032007 (2012); arXiv:1106.5685v2,
\bibitem{ref:mini}
  A. A. Aguilar-Arevalo {\it et al.} [MiniBooNE Collaboration], 
  Phys.\ Rev.\ Lett.\ {\bf 103}, 061802 (2009), arXiv:0903.2465 [hep-ex].
  
\end{thebibliography}
\end{document}